\documentclass[prl,floatfix,showpacs,amsmat,twocolumn]{revtex4}
\usepackage{amssymb}
\usepackage{amsmath}
\usepackage{graphicx}
\usepackage[colorlinks]{hyperref} 
\usepackage{mathptm}

\def\ket #1{\vert #1\rangle}
\def\bra #1{\langle #1\vert}
\def\ketbra #1 #2 {\langle #1\vert #2 \rangle}

\newcommand{\be}{\begin{eqnarray}}
\newcommand{\ee}{\end{eqnarray}}
\newcommand{\bea}{\begin{eqnarray}}
\newcommand{\eea}{\end{eqnarray}}
\newcommand{\bma}{\begin{subequations}}
\newcommand{\ema}{\end{subequations}}

\begin{document}

\title{%
  Creation, manipulation, and detection of Abelian and non-Abelian
  anyons in optical lattices}

\author{%
  M.~Aguado$^1$, G.~K.~Brennen$^2$, F.~Verstraete$^3$,
  J.~I.~Cirac$^1$}

\affiliation{%
  $1$ Max-Planck-Institut f\"ur Quantenoptik,
  Hans-Kopfermann-Str.~1, Garching, D-85748, Germany \\
  $2$ Centre for Quantum Information Science and Security, Macquarie
  University, 2109, NSW Australia \\
  $3$ Fakult\"at f\"ur Physik, Universit\"at Wien, Boltzmanngasse 5,
  A-1090 Wien, Austria.}


\date{\today}


\begin{abstract}
  Anyons are particle-like excitations of strongly correlated phases
  of matter with fractional statistics, characterized by nontrivial
  changes in the wavefunction, generalizing Bose and Fermi statistics,
  when two of them are interchanged.  This can be used to perform
  quantum computations \cite{Kitaev}.  We show how to simulate the
  creation and manipulation of Abelian and non-Abelian anyons in
  topological lattice models using trapped atoms in optical lattices.
  Our proposal, feasible with present technology, requires an ancilla
  particle which can undergo single particle gates, be moved close to
  each constituent of the lattice and undergo a simple quantum gate,
  and be detected.
\end{abstract}

\maketitle


The quest for physical systems where anyons \cite{WilczekOriginal} can
be observed has concentrated so far in effectively 2$d$ materials
exhibiting topological order \cite{WenOrder}.  Abelian anyons, whose
interchange generates a nontrivial phase in the wavefunction, exist in
the Fractional Quantum Hall effect.  Non-Abelian anyons, whose
interchange effects full unitary gates on the wavefunction, are
expected at certain filling fractions \cite{MooreRead} (see recent
experimental progress in \cite{Dolev}).  In spin lattice systems,
anyons can appear as low-lying excitations of topologically ordered
ground states (see, e.g., \cite{Kitaev, Levin, Fendley}).  Several
implementations of lattice models with anyonic excitations have been
put forward \cite{Doucot}\cite{xue}\cite{paredes}\cite{duan}%
\cite{micheli}\cite{Jiang}\cite{han}.  Those involving atoms or
molecules in optical lattices are especially attractive, given recent
experimental progress \cite{BlochRMP}.  Specifically, Kitaev's
honeycomb lattice model \cite{Kitaev:06} can be engineered
\cite{duan,micheli}, and anyonic interferometry in its Abelian phase
can be performed with cavity-mediated \emph{global} string operations
\cite{Jiang} or using individual addressing to braid excitations
\cite{ZhangProcNatl} (but due to the perturbative nature of the
effective Hamiltonian in this model, the visibility of anyonic
interferometry is degraded \cite{jvidal, Jiang}.)

Here we propose a novel scheme to create topologically ordered states,
generate and braid anyons, and detect their statistics for any setup
based on particles in optical lattices.  We use a lattice of particles
of species A to build the topological code and an ancilla of different
species B that can be moved independently and brought close to any A
particle to perform controlled operations on the code
\cite{auxspecies}.  Preparing the ancilla in superposition states,
making it interact with appropriate code particles and measuring its
state, the following can be achieved: \emph{creation of a topological
  state}, or a general error-correcting code (ECC); \emph{creation,
  braiding, and measurement (fusion) of anyons}, all operations needed
to perform topological quantum computation (TQC) by braiding; and
\emph{anyonic interferometry}, allowing direct observation of anyonic
statistics.  Note that (i) by using an ancilla with different quantum
states to perform the manipulation of the anyons, tasks can be carried
that are not possible using classical (e.g., laser) manipulation of
anyons (without single-particle addressability, all proposed methods
lack the power of ours); (ii) there is no need in principle of
single-particle addressability, specially to perform
proof-of-principle experiments; (iii) it is based on successfully
demonstrated technologies \cite{Porto:07}\cite{MeschedeSingleatom}%
\cite{BlochMott}; (iv) it is the first realistic protocol for
simulating \emph{universal} TQC in an atomic, molecular, optical
system system [while engineering the microscopic Hamiltonian to build
topological protection may be some time off (though see \cite{Bacon}),
the method herein works independently of the existence of the
background Hamiltonian.]

\begin{figure}
\centering
\includegraphics[clip=true]
  {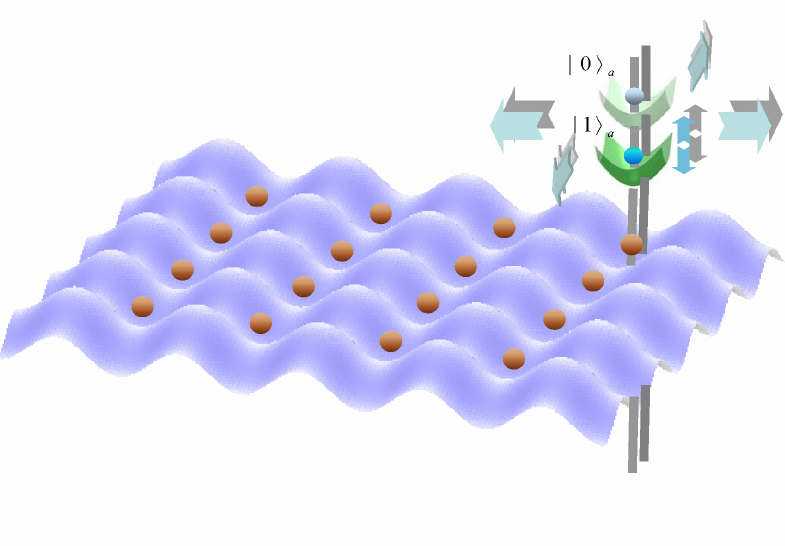} \\
  \caption{\label{fig1}%
    2$d$ optical lattice in the $x-y$ plane loaded with one atom per
    site.  Atoms are in different Zeeman levels $| 0 \rangle$ and $| 1
    \rangle$ in their ground electronic state, storing quantum
    information.  An ancillary atom of a different species in its
    electronic ground state, with relevant Zeeman levels $| 0
    \rangle_a$ and $| 1 \rangle_a$, is trapped using laser standing
    waves along the three directions.  Lasers in the $x-y$ plane are
    far detuned from the fine structure splitting: The potential
    controlling horizontal movement of the ancilla does not depend on
    its internal state.  The laser propagating along the $z$ direction
    is tuned in between the fine-splitted excited $P$ levels: The
    potential controlling the vertical movement depends on the
    internal state \cite{JakschColdcollisions, BrennenDipole}.}
\end{figure}

We consider 2$d$ lattices loaded with atoms or molecules, e.g.,
$^{87}$Rb.  The ancilla, e.g., $^{23}$Na, can be moved independently
using a laser potential not affecting Rb atoms (see Fig.~\ref{fig1}.)
These are now routinely loaded in optical lattices, and in the Mott
insulator state one can have extended regions with one particle per
site \cite{JakschMott, BlochMott}.  Single particles can also be
loaded in optical potentials and moved without decoherence
\cite{MeschedeSingleatom}.  Our scheme can be extended to layered 3$d$
configurations \cite{Dennis}.  We first consider the toric code
Hamiltonian \cite{Kitaev}, with Abelian anyonic excitations only, as a
toy model, but our scheme is basically model independent; later we
apply it to the $\mathrm{D} (S_3)$ quantum double model \cite{Kitaev},
which has non-Abelian anyons and is universal for TQC \cite{Mochon}.)

When the ancilla is brought close to a code atom, they experience a
2-qubit unitary $U_Z = | 0\rangle_a \langle 0 | \otimes I + | 1
\rangle_a \langle 1 | \otimes Z$ between their internal levels ($X$,
$Y$, $Z$ are Pauli operators.)  This gate can be implemented by cold
collisions \cite{JakschColdcollisions} or any other means
\cite{BrennenDipole, LukinRydberg}.  Single qubit operations can be
applied to the ancilla without having to address it, due to the
different level structure of code atoms.  Gates like $U_X = | 0
\rangle_a \langle 0 | \otimes I + | 1 \rangle_a \langle 1 | \otimes X$
can be implemented by applying appropriate gates before and after
$U_Z$.  The internal ancilla state (measurement of $Z_a$) can also be
detected with standard techniques without having to address it or
affect the code.

We next show how to create Kitaev's toric code \cite{Kitaev} with
these tools.  The code is defined as the ground level of a stabilizer
Hamiltonian on a square lattice of qubits, realized as Rb atoms, at
the edges of a square lattice.  The Hamiltonian $H = -
\sum_{\mathrm{v}} A_{\mathrm{v}} - \sum_{\mathrm{p}} B_{\mathrm{p}}$
is the sum of mutually commuting stabilizers $A_{\mathrm{v}} = \prod_{
  i \in \mathrm{v} } X_i$ and $B_{\rm p} = \prod_{ i \in \mathrm{p} }
Z_i$, where v runs over all vertices and p over plaquettes and
products involve the qubits surrounding the vertices or plaquettes.
We can associate the presence or absence of particle-like excitations
at the plaquettes (magnetic defects) and vertices (electric defects)
with the fulfillment or not of the ground level conditions
$A_{\mathrm{v}} = +1$ and $B_{\mathrm{p}} = +1$.  Plaquette and vertex
excitations are thus characterised by $B_{\mathrm{p}} = -1$ and
$A_{\mathrm{v}} = -1$ and appear at the ends of strings of $X$ and $Z$
operators applied on a ground state.  These `particles' turn out to
have nontrivial (anyonic) exchange statistics due to the
anticommutation of the $X$ and $Z$ Pauli operators, namely the
wavefunction gets multiplied by $-1$ when a vertex particle winds
around a region containing a single \nobreak{plaquette} excitation.
Detection of this phase change is possible via interference
experiments involving superpositions of states with and without
anyons.  Moreover, the degeneracy of the code allows to interpret it
in terms of a set of logical qubits whose $Z$ and $X$ operators are
given in terms of chains of $Z$'s and $X$'s.

\begin{figure}
\centering
\includegraphics
  {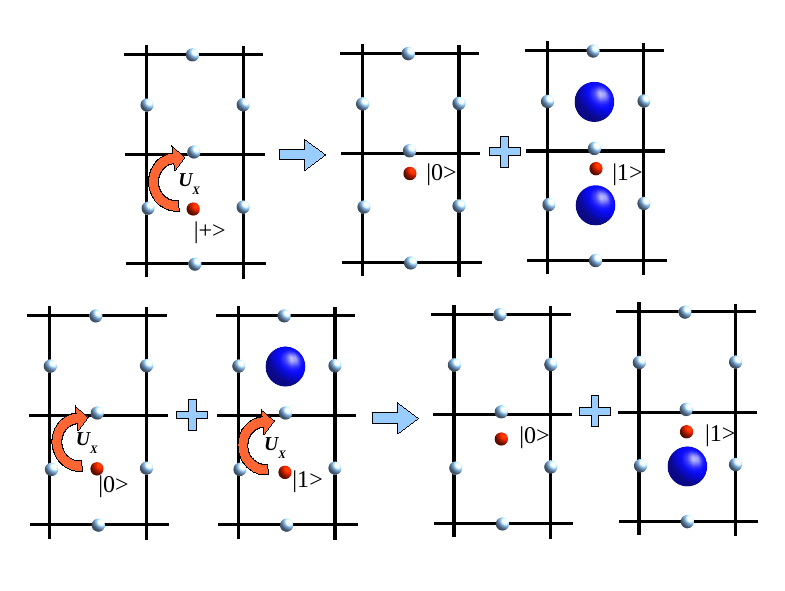} \\
  \caption{\label{fig2}%
    \emph{Top}: Creation of a superposition of the vacuum and a pair
    of magnetic defects.  On top of the ground state (code atoms shown
    as light circles), an ancilla (dark) is initialised as $\lvert +
    \rangle$ and brought close to a code atom, effecting $U_X = \lvert
    0 \rangle \langle 0 \rvert \otimes I + \lvert 1 \rangle \langle 1
    \rvert \otimes X$ and creating two magnetic defects (big blobs) in
    the adjacent plaquettes in the superposition component where the
    ancilla is in state $\lvert 1 \rangle$, while the code remains in
    the ground state in sector $\lvert 0 \rangle$.  \emph{Bottom}:
    Anyon transport.  The ancilla interacts via $U_X$ with a code atom
    between a plaquette satisfying the ground state condition and one
    of the plaquettes containing a magnetic defect in a superposition
    sector, transferring the anyon to the first plaquette.}
\end{figure}

We work with a rectangular surface with smooth and rough boundaries
\cite{Dennis}, with appropriate 3-body vertex and plaquette operators
along the boundary providing for a two-dimensional code space: One
logical qubit is encoded as the eigenvalue of a chain of $Z$'s along
an edge path connecting the rough boundaries.  The code space is
spanned by $+1$ coeigenstates of the stabilizers.  To create a state
$| \Psi \rangle$ in the code, we start with a well defined state $| 0
\rangle^{\otimes N}$ ($+1$ eigenstate of each $B_{\rm p}$) and measure
the $A$ stabilizers sequentially, from left to right and top to
bottom, using the ancilla.  If all outcomes are $+1$, our goal is
achieved, since $| \Psi \rangle \propto \prod_{{\rm v}} ( 1 + A_{\rm
  v} ) | 0 \rangle^{\otimes N}$ \cite{VerstraetePEPS}.  If $-1$ is
obtained, we can correct by applying $Z_b$ to qubit $b$ at the bottom
of the vertex using again the ancilla, since $Z_b ( 1 - A_{\rm v} ) |
0 \rangle^{\otimes N} = (1 + A_{\rm v} ) | 0 \rangle^{\otimes N}$;
$Z_b$ is applied to a qubit that has not been measured yet (for the
last row, $Z_b$ can be applied to the rightmost qubit.)  Once we have
measured all vertices and thus prepared the state, we can measure all
stabilizers to detect errors and apply error-correcting $X$'s or $Z$'s
to the corresponding qubits by using the ancilla.  We could have
started at another state, measured all operators in any order, and
then corrected errors in this way to prepare the desired state (this
can be used to prepare the target state in models beyond Kitaev's.)
We now show how plaquette and vertex measurements, as well as $X$'s
and $Z$'s, are performed using the ancilla.  To measure $A_{\rm v}$,
we prepare the ancilla in state $| + \rangle_a \propto | 0 \rangle_a +
| 1 \rangle_a$, move it to each qubit in the vertex, and apply $U_X$
each time.  Then we apply a Hadamard gate to the ancilla and measure
$X_a$.  If the result is $\pm 1$, we have applied $\langle \pm |
\prod_{ i \in {\rm v} } U_X | + \rangle = ( 1 \pm A_{\rm v} )$ to the
qubits at the vertex, thus performing the desired measurement.
$B_{\rm p}$ is measured by substituting $U_Z$ for $U_X$.  To apply $X$
($Z$) to a qubit, we prepare the ancilla in state $| 1 \rangle_a$,
approach it to that qubit and apply $U_X$ ($U_Z$.)  Once the toric
code state is prepared, operations within the code are performed by
applying strings of operators, using the ancilla in state $| 1
\rangle_a$, and applying $U_X$ or $U_Z$ sequentially on the desired
qubits by bringing them close to the ancilla.  To measure string
operators, prepare the ancilla in state $| + \rangle$, follow the same
sequence, and measure $X_a$ at the end.  The toric code has two kinds
of elementary excitations \cite{Kitaev}: pairs of anyons in frustrated
vertices (electric defects, $A_{\rm v} = -1$) and in frustrated
plaquettes (magnetic defects, $B_{\rm p} = -1$), with mutual Abelian
anyonic statistics.  They can be created by applying $Z$ or $X$ to a
given qubit, and can be moved, braided, and fused together by applying
these operators along a given path using the ancilla.  Superpositions
of states with and without vortices, or where they are in different
places (see Fig.~\ref{fig2}), can be created, allowing the observation
of fractional statistics: The simplest interference experiment is
shown in Fig.~\ref{fig3}.  On how to infer anyonic statistics from
interference experiments, see the Appendix.

\begin{figure}
\centering
\includegraphics
  {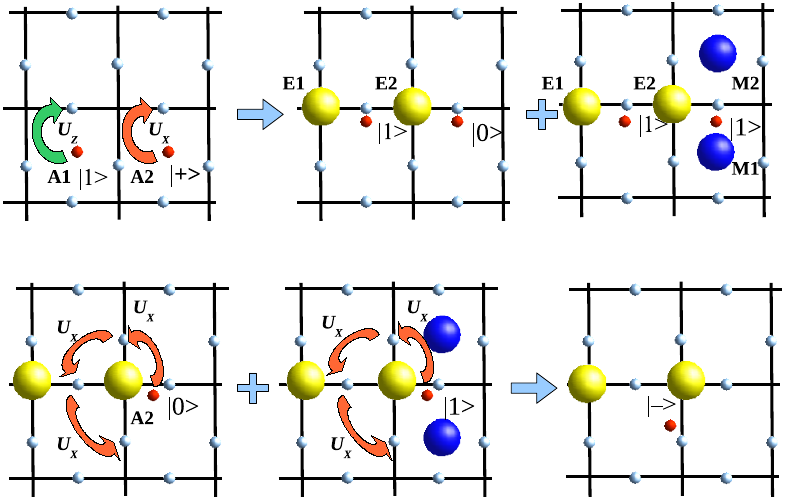} \\
  \caption{\label{fig3}%
    Minimal interferometry experiment for the toric code. \emph{Top}:
    Ancilla A1, initialised as $\lvert 1 \rangle$, unconditionally
    creates a pair E1, E2 of electric defects at neighboring vertices
    by application of $U_Z = \lvert 0 \rangle \langle 0 \rvert \otimes
    I + \lvert 1 \rangle \langle 1 \rvert \otimes Z$.  Ancilla A2,
    initialised as $\lvert + \rangle$, creates a superposition of the
    presence and the absence of two magnetic defects M1, M2 at
    neighboring plaquettes by application of $U_X$.  \emph{Bottom}: M1
    is wound around E2 by sequential $U_X$ interaction of ancilla A2
    with the code atoms surrounding E2.  M1 and M2 are eventually
    reannihilated, bringing both sectors to the ground state with a
    relative minus sign, $\lvert \text{GS} \rangle_{ \text{code}}
    \otimes \lvert 1 \rangle_{\mathrm{A}1} \otimes \lvert -
    \rangle_{\mathrm{A}2}$, i.e., a phase $-1$ is generated in the
    sector where braiding of defects takes place.  In this case, the
    interferometry results can be read from the ancilla lattice by a
    local projective measurement on A2.}
\end{figure}

\begin{figure}
\centering
\includegraphics[width=8cm]{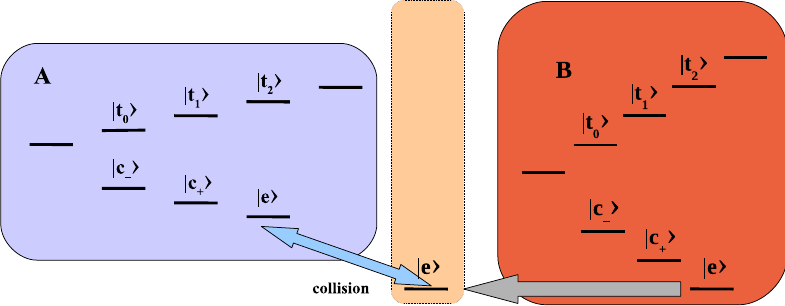} \\
\caption{\label{fig4}%
  Level structure for the $\mathrm{D} (\mathrm{S}_3)$ model.  The
  elements of $S_3$ are encoded into ground electronic hyperfine
  states of a trapped alkali atom A ($^{87}$Rb or $^{23}$Na, with $8$
  levels, or $^{133}$Cs, with $16$ levels.)  A 6-state ancilla $B$ of
  a different species is used to control operations.  Ancilla states
  can be moved independently: bringing $\lvert e \rangle_B$ close to a
  code atom A and coupling it to $\lvert e \rangle_A$, a collisional
  phase gate $\mathcal{Z}_e = I_A \otimes I_B - 2 \lvert e \rangle_B
  \langle e \rvert \otimes \lvert e \rangle_A \langle e \rvert$ is
  obtained; together with simultaneous 1-particle operations
  ${\mathcal{V}}_{\text{all A}} = \bigotimes_A v_A$ on the code, it
  provides all controlled operations.  Indeed,
  ${\mathcal{V}}_{\text{all A}}^\dagger \mathcal{Z}_e
  {\mathcal{V}}_{\text{all A}}$ yields controlled phase gates for any
  code state with $v_A = \lvert g \rangle_A \langle e \vert + \lvert e
  \rangle_A \langle g \vert$, and transpositions with $v_A \propto (
  \lvert e \rangle + \lvert g \rangle )_A \langle e \vert + ( \lvert e
  \rangle - \lvert g \rangle )_A \langle g \vert$ (these can be
  composed to obtain any controlled group multiplication.)}
\end{figure}

We now outline the preparation and manipulation of anyons in a
non-Abelian setting universal for quantum computation \cite{Mochon}:
the lattice quantum double model $\mathrm{D} ( S_3 )$ based on the
group of permutations of 3 elements $\mathrm{S}_3$ \cite{Kitaev} (see
a brief discussion in the Appendix; the full construction is given in
\cite{longversion}.)  It is a lattice model generalizing the toric
code, where local degrees of freedom live at the (oriented) edges of
the lattice with orthonormal bases $\{ \lvert g \rangle \}$ labeled by
the six group elements $g \in G$.  The Hamiltonian, also of the form
$H = - \sum_{\mathrm{v}} A_{\mathrm{v}} - \sum_{\mathrm{p}}
B_{\mathrm{p}}$, has commuting vertex and plaquette stabilizers
imposing constraints on the ground states.  Their violations define
particle-like excitations (anyons) with topological charges (electric,
magnetic and dyonic), with non-Abelian fusion and braiding rules.
Creation, transport, and fusion of anyons can be achieved generalizing
the controlled-NOT operations of the toric code to controlled left and
right group multiplications: $\mathcal{U}^{\mathrm{L}, \,
  \mathrm{R}}_h = \lvert 0 \rangle_B \langle 0 \rvert \otimes I_A +
\lvert 1 \rangle_B \langle 1 \rvert \otimes ( \sigma^{\mathrm{L}, \,
  \mathrm{R}}_h )_A$ with $\sigma^{\mathrm{L}}_h \lvert g \rangle =
\lvert h g \rangle$, $\sigma^{\mathrm{R}}_h \lvert g \rangle = \lvert
g h \rangle$.  The local degrees of freedom for the $\mathrm{D}
(\mathrm{S}_3)$ model are qudits of six dimensions, and their six
basis elements can be encoded into ground electronic hyperfine states
of an alkali atom with enough levels, used as code lattice A.  To
create, transport, and fuse pure electric and magnetic charge states
(enough to simulate universal TQC \cite{Mochon}), a 6-state ancilla
species B is especially appropriate (see Fig.~\ref{fig4}.)  As in the
toric code, vertex operators can be measured using ancilla-assisted
operations to prepare the ground state (see the Appendix.)  With one
single ancilla, which need not be spatially addressed, our algorithm
requires $\mathcal{O} ( n m )$ steps in an $n \times m$ region; with
an auxiliary lattice with one ancilla per face of the code lattice,
assuming addressability, it can be parallelized to depth $\mathcal{O}
( n + m )$ (essentially optimal \cite{Bravyi:06}; see the Appendix.)

This scheme is independent of the method used to construct the
topological state.  It can be built by cooling an atomic ensemble
interacting via an engineered topological Hamiltonian, providing in
principle topological protection to the code except for anyonic
manipulations, which should take the system to excited levels in a
controlled way (as needed to perform TQC as such; on how to simulate
relevant Hamiltonians, see \cite{duan, micheli}.)  But it can also be
constructed by the above procedure using ancillas to impose stabilizer
constraints, enough to perform proof-of-principle interference
experiments, a worthy goal by itself.  This also allows fault-tolerant
quantum computation with general ECCs (topological codes are excellent
ECCs: local operators do not mix topological sectors, string operators
mixing them can be efficiently implemented.)

The arrangement of ancillas is flexible.  One ancilla, individually
manipulable \cite{MeschedeSingleatom}, suffices in principle.  Varying
degrees of parallelization are possible: the ground state can be
constructed with sequential operations on one column of the sample
(parallel to smooth boundaries) at a time, so as to cover the whole
sample; in interferometry, commuting operations can be done
simultaneously; some parallelization can be introduced with a coarser
optical potential for the ancilla than for the code.

This method shares common problems of optical lattice schemes of
quantum computation, in particular, spontaneous emission.
Essentially, only the vertical direction of the ancilla is close to
resonance; lifetimes of seconds can be reached by tuning the laser
between the fine structure levels and can be enhanced by restricting
manipulation of the vertical dynamics to the (short) times the ancilla
is close to a code atom.  The ancilla can be repumped after an
operation, allowing one to repeat the tasks and detect errors.
Controlled logic by cold collisions \cite{JakschColdcollisions}
requires cooling the system to the physical ground state, possible for
both the code (in the 1st Bloch band, the ground state of the local
potential) and the ancilla.  Rydberg gates \cite{BrennenDipole,
  LukinRydberg} based on dipole-dipole interactions eliminate this
condition.  Ancilla-code interactions must break the code in a
controlled way, not creating (superpositions of) stray anyons spoiling
the quantum memory: Theoretical analysis \cite{Calarco} and
experimental results \cite{MeschedeRegister} suggest that excellent
control and small decoherence rates are achievable.  Then the
implementation benefits from the added protection of topological codes
(or, in general, ECCs.)  With a bias magnetic field, arbitrary single
qudit unitaries can be realized using frequency and polarization
selectivity of microwave or Raman laser pulses.  Collisional gates can
be realized using trap-induced shape resonances \cite{Stock:03}, or
using Raman pulses to map code and ancilla ground states
($\ket{e}_{A,B}$ in figure \ref{fig4}) to a vibrational excited state
of each lattice well, evolving by a collisional phase and mapping back
\cite{Strauch:07}.

The experimental techniques required by our method are (i) independent
trapping of two particle species A and B with different laser trapping
potentials; (ii) diluting the population of species B so that each
particle is individually addressable; (iii) bringing species A to a
Mott insulator phase; (iv) initialisation of species A in a product
state $\lvert 0 \rangle^{\otimes N}$; (v) single particle gates on
species B; (vi) simultaneous gates on all particles of species A;
(vii) independent transport of internal states of single particles of
species B so as to effect cold collisions with particles of species A.
Additionally, large scale simulations would require the ability to
recool to vibrational states.  Each one of these techniques has been
demonstrated experimentally; bringing them together will pose an
interesting experimental challenge.

A method based on the availability of an ancilla species thus allows
one to perform all TQC tasks on a given topological state,
independently of the way in which this state is constructed, in
optical lattices; e.g., universal TQC based on braiding can be
performed on top of the ground state of the $\mathrm{D} ( \mathrm{S}_3
)$ model.  This scheme is likely to prove the most practical and
general way to perform TQC in optical lattices.  Large computations
will face a steep scaling (a problem not exclusive of topological
settings), but observing interference phenomena and applying gates by
anyon braiding is feasible with today's technology.  This method can
also be used advantageously in general ECCs.

\textbf{Acknowledgements:} Financial support from the EU (SCALA) and
DFG (MAP and NIM Excellence Clusters) is gratefully acknowledged.
M.~A.~thanks S.~Iblisdir, D.~P\'erez-Garc\'{\i}a and J.~Pachos for
discussions on the quantum double models.

\appendix
\section{\label{appendix}%
  Appendix: Supplementary material}

In this Appendix we describe in some detail how to simulate
non-Abelian anyonic models.  Our construction is based on a
generalization of the toric code which is spin lattice realization of
the quantum double $D(G)$ of a discrete group $G$.  To realize this we
work on a connected two complex
$\Gamma=\{\mathcal{V},\mathcal{E},\mathcal{F}\}$ which is a
cellulation of a two dimensional surface described by vertex, edge,
and face sets.  Particles with $d=|G|$ levels (qudits) are placed on
the edges and physical states live in a Hilbert space
$\mathcal{H}=\mathcal{H}(d)^{\otimes |\mathcal{E}|}$ where
$\mathcal{H}(d)=\{\mathbb{C}\ket{j}\}_{j=0}^{d-1}$.  As in the toric
code, particles on edges that meet at a vertex $v$ all interact via a
vertex operator $A(v)$.  Similarly, all particles on edges that are on
the boundary of a face $f$ interact via $B(f)$.  We pick an
orientation for each edge with $e=[v_j,v_k]$ denoting an edge with
arrow pointing from vertex $v_j$ to $v_k$.  The choice of edge
orientations is not important as long as a consistent convention is
used.  The face orientations are all chosen consistent with the
surface orientation.  Whereas in the toric code model, the ground
states are invariant under local $\mathbb{Z}_2$ gauge transformations
generated by $\{A(v)\}$, here we demand gauge invariance under
transformations: \be T_g(v)=\prod_{e_j\in [v,*]}L_g(e_j)\prod_{e_j\in
  [*,v]}R_{g^{-1}}(e_j), \ee where $L_g(e_j),R_g(e_j)\in U(d)$, the
$d$ dimensional unitary group, are the permutation representations of
the left and right action of multiplication by the group element $g\in
G$ on the system particle located at edge $e_j$. For the particle
states we make the identification $\ket{j}\equiv\ket{g_j}$, where by
convention $\ket{0}\equiv\ket{g_0}\equiv\ket{e}$, with $e$ the
identity element. The action of left and right group multiplication on
the basis states is then $L_h\ket{j}=\ket{h g_j}$,
$R_h\ket{j}=\ket{g_j h}$.  A suitable spin lattice model was
introduced by Kitaev \cite{Kitaev}:
\begin{equation}
H_{\rm TO}=-\sum_v A(v)-\sum_f B(f)
\end{equation}
where 
\begin{equation}
\begin{array}{lll}
  A(v) = \displaystyle\frac{1}{|G|}\sum_{g\in G}T_g(v)&&, \\
  B_{\ell}(v,f) = \displaystyle
    \sum_{\{
       \prod_{ e_k\in\partial f} g_k^{-o_f(e_k)}=\ell | e_0=[v,*]
    \}}
    \otimes_{e_k\in \partial f}
    \ket{g_k^{-o_f(e_k)}}_{e_k}\bra{g_k^{-o_f(e_k)}}&&,
\end{array}
\end{equation}
The operator $A(v)$ is a projector onto gauge symmetrized states.
Similarly, the operator $B_{\ell}(v,f)$, is a projector onto states
with flux $\ell$ at face $f$.  Here the sum is taken over all products
of group elements $g_k$ acting on a counterclockwise cycle of edges on
the boundary of $f$ such that the product is $\ell \in G$.  The
function $o_f(e_j)=\pm 1$ according to whether the orientation of the
edge is the same as (opposite to) the face orientation.  Assigning the
origin $v$ from which the product is taken is important as failure to
do so is equivalent to only specifying the conjugacy class or magnetic
charge.  Since the trivial flux is its own conjugacy class we write it
$B_e(v,f)=B(f)$, which is the projection we seek for ground states of
$H_{TO}$.

By construction $[A(v),A(v')]=[B(f),B(f')]=[A(v),B(f)]=0$.  The ground
states of $H_{\rm TO}$ are then manifestly gauge invariant states
defined as $+1$ coeigenstates of $\{A(v),B(f)\}$.  Excited states are
described by violations of the local constraints $A(v),B(f)$ and are
particle-like corresponding to irreducible representations (irreps)
$\Pi^{[\alpha]}_{R(N_{[\alpha]})}$ where $[\alpha]$ labels the
magnetic charge denoting a conjugacy class of $G$, and
$R(N_{[\alpha]})$ labels the electric charge which denotes a unitary
irrep $R$ of the normalizer of an element in the conjugacy class
$[\alpha]$.  Notice that for the group $\mathbb{Z}_2$, $L_e=R_e={\bf
  1}_2$ and $L_{g_1}=R_{g_1}=\sigma^x$, and $H_{TO}$ is precisely the
toric code Hamiltonian.

Henceforth, we fix $G = S_3 = \{ g_j \}_{ j=0 }^5$ the symmetric group
of three objects and let $\Gamma$ be an $n\times m$ square lattice
with boundary.  We place $d=6$ qudits on edges labeling each by it's
location $e_{i,j;k,l}=[v_{i,j},v_{k,l}]$ and $d=6$ vertex and face
ancillas labeled by their locations $v_{j,k}$ and $f_{j,k}$.  For a
two complex with boundary the ground state $\ket{GS}$ of $H_{TO}$ is
unique \cite{Bullock:07} and can be constructed by measuring the
vertex projectors in an analogous way to the toric code.  We begin
with all system particles in state $\ket{e}=\ket{0}$ which satisfies
the zero flux condition $\langle B(f)\rangle=1\forall f$.  All vertex
and face ancillae are prepared in the state $\ket{\tilde{0}}$ where
$\ket{\tilde{j}}=\frac{1}{\sqrt{6}}\sum_{k=0}^{5} e^{2\pi ij
  k/6}\ket{k}$.  We then apply a sequence of operations columnwise
beginning on the left column of vertices $\{v_{k,0}\}$ and ending on
the right.  For column $k$ we apply the controlled operation
$W(v)=\sum_{h\in S_3}\ket{h}_{v}\bra{h}\otimes T_g(v)$ between each
vertex ancilla $v_{j,k}$ and its edge neighbors.  These operations can
be done in parallel since $[L_h(e),R_{h'}(e)]=0$.  Next we measure
each vertex ancilla $v_{j,k}$ in the basis $\{\ket{\tilde{r}}\}$ and
given the outcome $m(j,k)$ apply the single qudit correction gate
$Z^{m(j,k)}(e_{j,k;j,k+1})$, where $Z(e)=\sum_{k=0}^{5}e^{2i\pi
  k/6}\ket{k}_e\bra{k}$, on the right edge.  Finally when we reach the
rightmost column of vertices $\{v_{j,m-1}\}$ we apply a sequence of
operations from bottom to top involving applying $W(v_{j,m-1})$
followed by measurement of the ancilla and the correction gate
$Z^{m(j,m-1)}(e_{j,k;j,k+1})$ on the top edge.  When we reach vertex
$v_{n-1,m-1}$ we are done since the operator $T(v_{n-1,m-1})$ is not
independent but can be written as a product of the others meaning that
$\langle A(v_{n-1,m-1})\rangle=1$ already.
 
Our algorithm has a computational depth of $O(m+n)$ as measured by the
number of parallel elementary two qudit gates.  We find contructions
of the operators $\sum_{h\in S_3}\ket{h}\bra{h}\otimes L_h$ or
$\sum_{h\in S_3}\ket{h}\bra{h}\otimes R_h$ in $37$ controlled phase
gates $e^{i\pi \ket{0}\bra{0}\otimes \ket{0}\bra{0}}$ and roughly
twice that number of local pairwise basis state couplings and suspect
this count is optimal.  One might wonder if a faster ground state
preparation procedure is possible.  The answer is no if the initial
state is uncorrelated and the available set of operations is
quasi-local.  The reason is that the final state has global
correlations that are created by quasi-local operations.  In our
algorithm these operations are measurements but they could also be
adiabatic turn on of the summands of $H_{TO}$. The time scale to
perform the quasi-local operations (here the measurement time of
$A(v)$) establishes a light cone for the flow of correlations.  In
\cite{Bravyi:06} it was shown by an application of the Lieb-Robinson
bound that the minimal time to prepare a topologically ordered state
beginning in a completely uncorrelated state is of the order of the
length of the correlations.  Since the correlation length scales as
the linear dimension of our lattice, our algorithm is essentially
optimal.

The group $S_3$ has three conjugacy classes: the identity $[e]=\{e\}$,
transpositions $[t]=\{t_0,t_1,t_2\}$ and cyclic permutations
$[c]=\{c_+,c_-\}$, and three irreps: the two one dimensional irreps
$R_1^+(g) = 1$, the signature representation
\[
\label{sthree:eq:signature}
  R_1^- ( e )   = +1 = R_1^- ( c_\rho ) ,
\quad
  R_1^- ( t_i ) =   -1 ,
\]
and the two-dimensional irrep
\[
\nonumber
  R_2 ( e )
=
  \mathbf{1}_2 ,
\quad
  R_2 ( t_k )
=
  \sigma^x
  e^{i \frac{2\pi}{3} \, k \, \sigma^z},
  R_2 ( c_\rho )
=
  e^{ i\, \rho\, \frac{ 2 \pi }{ 3 } \, \sigma^z}.
\]

For $D(S_3)$ there are $8$ irreps: the vacuum state
$\Pi^{[e]}_{R_1^+}$, pure magnetic charges $\Pi^{[c]}_{\beta_0},
\Pi^{[t]}_{\gamma_0}$, pure electric charges $\Pi^{[e]}_{R_1^-},
\Pi^{[e]}_{R_2}$, and dyonic combinations $\Pi^{[c]}_{\beta_1},
\Pi^{[c]}_{\beta_2},\Pi^{[t]}_{\gamma_1}$.  A complete derivation of
the fusion rules and braid relations for this model is given in
\cite{Propitius:95}.  We focus here pure electric or pure magnetic
charge states, which in fact are sufficient for universal topological
quantum computation \cite{Mochon}.

Pure electric charges are labeled by basis states
$\ket{(P^{R}_{\mu,\nu},e);(v,-)}$ where $P_{\mu, \nu}^R =
\frac{|R|}{|G|} \sum_{g\in G} [R(g)^*]_{\mu,\nu} g$ is the projection
operator onto a subspace belonging to the unirep $R$.  Since we are
considering pure electric charges, the unireps of the normalizer of
$[e]$ are equivalent to the unireps of $G$ itself.  In the context of
the spin lattice model this is interpreted as electric charge created
by applying the projection operator $P^R_{\mu,\nu}$ onto the system
with the group action being local gauge transformations $T_g(v)$.
Magnetic fluxes are labeled by basis states:
$\ket{(P^{R_1^+},\ell);(v,f)}$ and are understood as the result of a
projection $B_{\ell}(v,f)$.

Excitations created in the bulk always come in particle anti-particle
pairs.  A generic state of a magnetic charge pair $([\ell, \,
\ell^{-1}])$ is
\[
  \sum_{\ell\in[\ell]}
  c_{\ell}
  \ket{(P^{R_1^+},\ell^{-1});(v_{i,j},f_{i,j})}
  \ket{(P^{R_1^+},\ell);(v_{i,j},f_{i,j+1})}
\]
with $\sum_{\ell\in [\ell]}|c_{\ell}|^2=1$.  The unique ${\it vacuum}$
magnetic charge pair state invariant under conjugation by fluxes is
the state with $c_{\ell}=1/\sqrt{|[\ell]|}\forall \ell$, denoted
$\ket{0_{[\ell]};(v,f),(v',f')}$.  Note the vacuum state with
neighboring magnetic charge pairs can be written
\begin{equation}
  \ket{0_{[\ell]};(v_{i,j},f_{i,j}),(v_{i,j},f_{i,j+1})}
 =
  \frac{1}{\sqrt{|[\ell]|}}
  \sum_{\ell\in[\ell]}R_{\ell}(e_{i-1,j;i,j}) \ket{GS}
\end{equation}
We can prepare this by beginning with the ancilla $f_{i,j}$ in the
state $\ket{0_{[\ell ]}}_{f_{i,j}}$ where $\{ \ket{k_{[\ell]}} =
Z_{|[\ell]|}^k \ket{0_{[\ell]}} \}_{k=0}^{|[\ell]|-1}$, with
$Z_{[\ell]}^k = \sum_{\ell_m\in[\ell]} e^{i2\pi k m/|[\ell]|}
\ket{\ell_m}\bra{\ell_m}$ and we have labeled the group elements in
$[\ell] = \{l_0,\ldots, \ell_{|[\ell]|-1}\}$), and applying the two
qudit unitary
\begin{equation}
  F_{[\ell]}(f_{i,j})
=
  {\bf 1}_{|G|-|[\ell]|} \otimes {\bf 1}_{|G|}
 +
  \sum_{\ell\in [\ell]}
  \ket{\ell}_{f_{i,j}}\bra{\ell} \otimes R_{\ell}(e_{i-1,j;i,j})
\end{equation}
Next we measure the face ancilla in the basis $\{\ket{k_{[\ell]}}$.
For the outcome $0_{[\ell]}$ the target magnetic charge state is
created. Otherwise for outcome $k_{[\ell]}$, we need a correction
step.  To do this prepare the ancilla $f_{i,j}$ in the state
$\ket{e}_{f_{i,j}}$ and apply the controlled operation
$\Lambda(v_{i,j},f_{i,j})$ where
\begin{equation}
  \Lambda(v,f)= \sum_{g\in G}B_{g}(v,f)\otimes L_{g}(f).
\end{equation}
which maps the ancilla $f$ to state $\ket{g}_f$ when the flux at $f$
evaluated with base point $v$ is $g$.  Such a controlled operation can
be decomposed into elementary two qudit controlled rotation operators
with each edge $e_k$ surrounding $f$ as a control and the ancilla as
the target.

In the toric code and in fact for discrete gauge theories for all
finite Abelian groups \cite{Bullock:07}, excitations can be propagated
by applying one local operator which simultaneously annihilates a
charge at one location and create one at a neighboring location.  This
is {\it not} true for non-Abelian theories as doing so violates a face
or vertex constraint. To propagate magnetic charges from one face $f$
to an adjacent face $f'$ essentially involves coherently mapping the
value of flux at $f$ to the face ancilla $f$ using $\Lambda(v,f)$ and
applying a controlled operation on the shared edge of the faces
$f,f'$.  After this controlled operation the face ancilla $f$ is
disentangled from the system by mapping the flux at face $f'$ to
ancilla $f'$ and performing a controlled operation between ancillae
$f,f'$ and finally reversing the mapping on $f'$.  In this protocol we
are careful to demand only single qudit and nearest neighbor two qudit
interactions.  This entire process respects superpositions over flux
states and can therefore be used to propagate magnetic charges around
the lattice \cite{longversion}.  Fusion of a magnetic charge pair can
be measured by using controlled operations to bring the constituent
charges in conjugacy class $[\ell]$ adjacent to one another at faces
$(f,f')$ and applying the operator $\Lambda(v,f)= \sum_{g\in
  G}B_{g}(v,f)\otimes L_{g}(f)$ followed by measurement of the ancilla
$f$ in the basis $\ket{0_{[\ell ]}}_{f_{i,j}}$.  The probability to
obtain the outcome $0_{[\ell]}$ equals the probability for the pair to
fuse into the vacuum.

A generic state of an electric charge pair $(R, \, R^\ast)$ at
vertices $(v,v')$ will be represented as an $|R|\times |R|$ matrix
\begin{align}
\nonumber
 \ket{M^R;(v,v')}
&=
 \frac{1}{\sqrt{|R|}}
 \sum_{\mu,\nu} M^R_{\mu,\nu} \\
\nonumber
&\times
 \frac{ 1 }{ \sqrt{ \lvert R \rvert } }
 \sum_{ \beta = 0 }^{ \lvert R \rvert - 1}
 \ket{(P^R_{\mu,\beta},e);(v,-)} \ket{(P^{R*}_{\nu,\beta},e);(v',-)}
\end{align}
with $\sum_{\mu,\nu=0}^{|R|-1}|M^R_{\mu,\nu}|^2=|R|$. There is a
unique ${\it vacuum}$ electric charge state which is invariant under
conjugation by fluxes: $ \ket{{\bf 1}^{R};(v,v')}$.  The state
$\ket{{\bf 1}^{R};(v_{i,j},v_{i,j+1})}$ can be prepared as follows.
First prepare the vertex ancilla $v_{i,j}$ in state
$\ket{e}_{v_{i,j}}$. and apply the conditional unitary
$K(v_{i,j},e_{i,j;i,j+1})$ defined by
\begin{equation}
  K(v,e)
 =
  \left\{\begin{array}{c}
   \sum_{g\in G} \ket{g}_{e} \bra{g} \otimes R_{g}(v)
    \quad e=[v,*]  \\
   \sum_{g\in G} \ket{g}_{e} \bra{g} \otimes R_{g^{-1}}(v)
    \quad e=[*,v]
  \end{array}\right.
\end{equation}
Depending on the representation, $R$, apply a single qubit operation
$W_R$ on the ancilla $(v_{i,j})$:
\[
\begin{array}{lll}
 W_{R_1^+}={\bf 1}_6 &&\\
 W_{R_1^-}
=
 \ket{e}\bra{e} + \ket{c_+}\bra{c_+} + \ket{c_-}\bra{c_-}
 - \ket{t_0}\bra{t_0} - \ket{t_1}\bra{t_1} - \ket{t_2}\bra{t_2} &&\\
 W_{R_2}
=
 2 \ket{e}\bra{e} - \ket{c_+}\bra{c_+} - \ket{c_-}\bra{c_-}\\
\end{array}
\]
(the latter is not unitary but can be constructed using adaptive
measurements as detailed in \cite{longversion}) and finally apply
$K^{-1}(v_{i,j},e_{i,j;i,j+1})$ to disentangle the ancilla from the
system.  One can prepare more distant vacuum charge pairs,
e.g. $\ket{{\bf 1}^{R};(v_{i,j},v_{i',j'})}$ using the same procedure
but after each application of $K(v,e)$, swapping the vertex qudit with
the next relevant vertex qudit on the path (via an intermediary swap
with the in between edge qudit) then applying the operator
$U_R(v_{i',j'})$ at the end, and finally inverting the steps targeting
the ancilla.  Fusion of electric charge pair $\ket{M^R;(v,v')}$ is
measured by beginning at vertex $v$ and applying a sequence of
controlled operations $K(v,e)$ acting on a path of vertices from $v$
to $v'$.  At vertex $v'$, the operator $L_{c_+}(v')$ is applied and
then the set of controlled operations along the vertex path is
inverted.  Finally, the first ancilla $v$ is measured in any unitary
extension of the basis $\{\ket{R} = U_R \sum_{g\in S_3}
\ket{g}/\sqrt{6}\}$.  The probability to measure the outcome state
$\ket{R_1^{\pm}}$ is the probability for fusion in the vacuum or into
the signed irrep.

We now have all the steps to create and braid and fuse anyons.  One
can verify that all the usual rules for exchanging and braiding anyons
are satisfied \cite{longversion}.  In particular it is easy to see
that braiding charges around each other acts trivially since the
electric charge pair creation operators are diagonal in the logical
basis.  Furthermore, the gauge transformations $T_h(v)$ can be viewed
as creating a magnetic flux $h,h^{-1}$ pair, braiding them around the
vertex $v$ and annihilating.  To verify non trivial braiding, one can
look for imperfect fusion of electric or magnetic charge pairs into
the vacuum.  Let's describe a simplified interference experiment that
could be performed in principle on a single face of the lattice.  We
begin by preparing the state of adjacent electric charge pairs
$\ket{{\bf 1}^{R_2};(v_{i,j},v_{i,j+1})}$ and the vertex ancilla
$v_{i,j}$ in $\ket{h^+}$ where $\ket{h^{\pm}} = (\ket{e}\pm\ket{h}) /
\sqrt{2}$ for some $h\in S_3$.  Next we apply the controlled operation
$W(v_{i,j})$ which creates the state $(\ket{e}_{v_{i,j}} \ket{{\bf
    1}^{R_2};(v_{i,j},v_{i,j+1})} + \ket{h}_{v_{i,j}} \ket{{\bf
    1}^{R_2};(v_{i,j},v_{i,j+1})}) / \sqrt{2}$ followed by measurement
of the ancilla in the basis $\ket{h^{\pm}}$ with outcome $m=\pm 1$.
The outcome distribution satisfies
\[
\begin{array}{lll}
  P(m=1)-P(m=-1)
&=&
  |\bra{{\bf 1}^{R_2};(v,v')} R_2(h);(v,v')\rangle|^2\\
&=&
  \frac{|\mbox{Tr} [R_2(h)]|^2}{|R_2|^2}
\end{array}
\]
which is precisely the fusion probability for $R_2(h)\rightarrow {\bf
  1}^{R_2}$.

It is important to clearly define what we mean by inferring anyonic
statistics from interference measurements.  For any physical theory,
the phase accumulated when braiding one particle around another has
contributions from the statistics $\phi_s$ as well as possible
dynamical $\phi_d$ and geometric (Berry's phase) contributions
$\phi_g$ \cite{Levin:03}.  Considering the interferometry experiment
depicted in Fig.~\ref{fig3}, including the other phases in the process
gives the transformations $ \ket{GS}\ket{1}_{A1}\ket{+}_{A2}
\rightarrow \ket{E1,E2;(v_1,v_2)}\ket{1}_{A1}\ket{0}_{A_2} +
\ket{E1,E2}\ket{M1,M2}\ket{1}_{A_1}\ket{1}_{A_2} \rightarrow
\ket{E1,E2}\ket{1}_{A1}\ket{0}_{A_2} + e^{i(\phi_s+\phi_g+\phi_d)}
\ket{E1,E2}\ket{M1,M2}\ket{1}_{A_1}\ket{1}_{A_2}\rightarrow
\ket{GS}\ket{1}_{A1}(\ket{0}_{A2} +
e^{\i(\phi_s+\phi_g+\phi_d)}\ket{1}_{A2})/\sqrt{2}$.  If, e.g., there
were a background Hamiltonian $H=-U\sum_vA_v-U\sum_pB_p$ present, then
$\phi_d=t4U$ where $t$ is the time to perform the braiding.  The
statistical phase can be extracted by comparing this measured phase to
a proper {\it reference phase}. To do so we can perform a similar set
of operations but starting in the state
$\ket{GS}\ket{1}_{A0}\ket{+}_{A2}$ such that the created electric
defects neighboring A0 are not enclosed by the braiding path of the
magnetic defects.  Since the braiding paths and particle numbers in
each component are the same in both experiments, the difference in the
measured phases accumulated on A2 is $\phi_s$.

Our scheme is flexible in how the ancillary particles are introduced
and one could work entirely within a single two dimensional lattice
with in place face and vertex ancilla. For example, the retroreflected
square lattice of \cite{Porto:07} provides a bipartite square lattice
with addressable systems and ancillary particles.  By dynamically
changing the trapping fields strengths and phases, the face and vertex
ancilla can be moved along any cardinal direction to interact pairwise
with the system particles on the edges.  However, to build the ground
states and manipulate anyons, individual addressability of the
ancillary particles using, e.g., the techniques in \cite{Cho:07,
  Gorshkov:07} seems necessary.



\end{document}